\documentclass{article}
\def \inbar{\vrule height1.5ex width.4pt depth0pt}
\def \C{\relax\hbox{\kern.25em$\inbar\kern-.3em{\rm C}$}}
\def \R{\relax{\rm I\kern-.18em R}}
\def \ids{ {\cal I}_2}
\def \idf{{\cal I}_1}

\def \Hi{{\cal H}}

\def \BO{{\cal B}}

\def \Di{D^{I}_{1}}

\newcommand{\beq}{\begin{equation}}
\newcommand{\eeq}{\end{equation}}
\newcommand{\bea}{\begin{eqnarray}}
\newcommand{\eea}{\end{eqnarray}}
\newcommand{\nn}{\nonumber}

\newcommand{\Tr}{\hbox{Tr}}
\newcommand{\Str}{\hbox{Str}}

\begin{document}
\author{Osman Teoman Turgut${\,}$\\
 \\
${}^{}\,$Department of Physics, Bogazici University \\
80815 Bebek, Istanbul, Turkey\\ 
and\\
Feza Gursey Institute\\
Kandilli 81220, Istanbul, Turkey\\
  turgutte@boun.edu.tr}

\title{\bf Geometric Quantization on the Superdisc }
\maketitle
\large
\begin{abstract}
\large
 In this article we discuss the geometric quantization on a certain type of infinite
dimensional super-disc. Such systems are quite natural when we analyze coupled bosons and 
fermions. The large-$N$ limit of a system like that  corresponds to a certain  
super-homogeneous space.
First,  we define an example of a super-homogeneous 
manifold: a super-disc.  We  show that it  has a natural symplectic form,
it  can be used to introduce classical  dynamics once a Hamiltonian is chosen.
Existence of moment maps provide a Poisson realization of the underlying symmetry 
super-group. These are the natural operators to quantize via methods of geometric 
quantization, and we show that this can be done.
\end{abstract}
\large
\section{Introduction}
 Geometric quantization
is an interesting and useful program for  quantizing systems whose phase spaces  have a 
simple geometric description \cite{hurt}.
It  is not always the case that the 
phase space has a nice geometric structure,  and even if it does, 
the result of quantizing via this method does not actually 
solve the problem but in many cases just helps one to formulate it.
The geometric approach to quantization goes back to works of 
Berezin \cite{berezin2, berezin3, berezin4, berezin}
 and at about the same time appeared in the mathematics literature as well.

In this work we will extend our previous analysis \cite{2dgeo} 
to the context of super-geometry. This is interesting in two ways, one is purely mathematical,
it gives a natural way to construct unitary representations of the underlying symmetry group.
The other one is the possibility of understanding physical systems which have  coupled 
bosons and  fermions.
Super-geometry sets the natural arena for formulating and  studying  these problems.
Our approach  originates from ideas of Rajeev on the large-$N$ limit of 
field theories. Rajeev has shown that a proper large-$N$ limit of QCD in two 
dimensions has a natural phase space given by an infinite dimensional Grassmannian 
\cite{2dqhd}. This general philosophy can be extended to other cases
\cite{2dgeo, tolyateo}.
Whenever there is a mixture of fermions and bosons, the large-$N$ phase space is 
expected to be a certain kind of super-homogeneous manifold. In gauge theory,
we have shown that this space is given by 
a certain kind of super-Grassmannian \cite{tolyateo}.
If instead we are looking at a fermionic system which has only a finite number of degrees
of freedom coupled to a bosonic field theory, its large-$N$ limit can be formulated  
as a  certain type of super-disc.
This can be seen as follows:
we get for such a   system, in the language of creation and 
annihilation operators, bilinears of the form 
\bea
      N(p,q)={2 \over N} : a^{\dag\alpha}(p)a(q)_\alpha:, \  \  
M_{ij}={2\over N}\chi^{\dag \alpha }_i\chi_{\alpha j}, \  
  Q_i(p)={2\over N}\chi^{\dag\alpha}_i a_\alpha(p), \ \ 
\bar Q_i(p)={2\over N}a^{\dag\alpha}(p)\chi_{\alpha i}
,\eea
where we have a normal ordering $:\  :$, for only the  bosonic products and
$\alpha$ denotes a ``color" index.
These operators are the natural ones for the 
large-$N$ phase space of the theory. In general 
it may not be possible to express all the dynamical variables in terms of these 
bilinears, but if we restrict ourselves to the ``color invariant'' sector,
these are the only ones we can compose.
We note that this statement is strictly true when we look at a gauge 
theory in $1+1$ dimensions \cite{tolyateo} but for that  we need  infinite degrees of freedom 
for the fermions, and that requires an analog of the 
Grassmannian. In some other cases this is only an approximation to the full model,
 the validity of which has to be tested depending on the specifics.
As an example we write  down a non-relativistic model, where a bosonic self-coupled field also 
couples with localized fermionic sources,
\beq
      H=\int \Big(:\nabla \phi^{\dag\alpha} . \nabla \phi_\alpha  +m^2 \phi^{\dag\alpha} 
\phi_\alpha+ {\lambda^2\over 2}(\phi^{\dag\alpha}\phi_\alpha)^2 :  +  g \sum_i \rho(x)
(\phi_\alpha\chi^{\dag\alpha}_i(t)+ \phi^{\dag \alpha}\chi_{\alpha i}(t))\Big)
\eeq
These  models may exhibit rather nontrivial dynamics, depending on the dimension we may 
need to renormalize the coupling constants. Our approach with Hilbert-Schmidt operators 
excludes cases which require renormalization, although a general super-disc is still 
present.
The above operators actually provide a realization of the super-Lie algebra
$U({\cal H}^e_-,{\cal H}^e_+|{\cal H}_+^o)$ as we will see.
In fact, one can see that many super-Lie algebras have natural realizations by 
fermionic and bosonic operators\cite{bars}.

In this article we will only deal with the mathematical aspects of this problem and 
think of  geometric quantization as a method for constructing the 
quantum Hilbert space where the dynamics takes place.
Solving a specific  model perhaps should  be done first in the classical setting of 
the large-$N$ limit.

\section{The Superdisc} 
In this section we present a brief definition of the superdisc which we denote by 
$D^{I}_1$ following  the reference \cite{lesniewski}, we mostly adopt  their conventions.
As we will see there is a small difference between our approach and this
reference.
In the same  reference there is a nice discussion of other cases,
 which one can generalize in the same way,
but we choose to look at the above simpler case  for the sake of clarity. 
The previous paper by the same authors \cite{lesniewski2} give a more detailed discussion of the 
$U(1,1|1)$ case, since the general case in \cite{lesniewski} 
is treated in a  sussinct manner,
we prefer to give a detailed discussion and believe that some of the
explicit formulae could be useful for the reader.
The physically interesting case requires an additional complication compared to  the 
one in \cite{lesniewski}, one should look at an infinite
Grassmann algebra. We will briefly discuss this generalization, yet the results are not so 
simple and as rigorous as in the finite dimensional one.
Some other useful sources are the lectures of Kostant \cite{kostant} and 
the books by Berezin \cite{berezin1} and Manin \cite{manin}.

Let us consider two Hilbert spaces, $\Hi^e$ and $\Hi^o$
which correspond to the even and odd spaces respectively. In physically interesting cases
they are either both separable infinite dimensional, or the 
even one  is separable infinite dimensional and the odd one is finite dimensional.
To keep the rigor we will only deal with $\Hi^o_+$ finite dimensional,
but arbitrarily large. Let us assume that its dimension is 
$N$, later on we will extend this to infinite dimensions.
We will split the even space into positive and negative parts,
each piece being infinite dimensional, 
$\Hi^e=\Hi^e_-\oplus \Hi^e_+$. We will really think of the odd part as 
the positive subspace and denote it as $\Hi^o_+$, this is just for 
convenience at the moment since we have not attached any physical significance 
to $\Di$. 

We may denote the ${\bf Z}$ graded super-space as $\Hi$, which splits 
with respect to ${\bf Z}_2$ grading as 
${\cal H}^e|{\cal H}^o$. It will be  
better to decompose  this space as $\Hi=\Hi^e_-\oplus\Hi^e_+|\Hi^o_+$.
Let us introduce 
the set of complex super matrices $Z$ such that 
\beq 
 Z=[ w  \quad \theta]
\eeq
where 
$w: \Hi^e_+\to \Hi^e_-$ and $ \theta: \Hi^o_+\to \Hi^e_-$,
furthermore we require the following convergence conditions
$w \in \ids$ and $\theta  \in \ids$, where $\ids$ denotes the Hilbert-Schmidt ideal
in this context.
A super space is given by the algebra of smooth  functions living on it,
in any given super-chart ${\cal U}$ we have 
$C^\infty({\cal U})\approx C^\infty(U) \otimes \wedge({\bf C}^s)$
for some $s$, and here $U$ denotes the corresponding open set for the 
base manifold.
(In \cite{lesniewski} the underlying function algebra for the odd generators is chosen to be 
$\wedge ({\bf C}^{mq})$.  We will instead take the set of 
generators  as $\wedge ({\bf C}^n)$, and 
$\theta$ denotes the matrix of linear transformations  from  the super vector space
${\cal H}^o_+$ to ${\cal H}^e_-$).

Let us explain the meaning of these convergence conditions:
if we expand the matrix $w$ into a series 
\beq
    w=w_B+w_{a_1a_2}\xi^{a_1}\xi^{a_2}+...
\eeq
where $\xi^a$ denotes half of the odd generators and this series terminates. 
There are also    
hermitian conjugates, that is we have a set of coordinates 
$\xi^a$ and $\xi^{*a}$.(Since the base manifold is contractable this expression is true, 
otherwise we need to assume it on any given chart). 
Then, we assume
that each one of these matrices are in the Hilbert-Schmidt class, i.e.
$w_B^\dag w_B, w_{a_1a_2}^\dag w_{a_1a_2},$ ...,$w_{12...r}^\dag w_{12...r} \in {\cal I}_1$,
here we use ${\cal I}_1$ to denote trace class operators.
 This decomposition is basis dependent, but the condition is
basis invariant. It is possible to see this by looking at a change of basis 
which is given by an invertible super-matrix(non-type changing one):
\bea
   (SwS^{-1})_B&=&S_Bw_BS^{-1}_B ....\nn\cr
(SwS^{-1})_{a_1a_2...a_{2k}}&=&
  S_Bw_{a_1a_2...a_{2k}}S^{-1}_B+...+S_{a_1...a_{2m}}w_{a_{2m+1}...a_{2n}}
S^{-1}_{a_{2n+1}...a_{2k}}...
\eea
etc, and we see that each component is replaced by a sum, each term of which 
is conjugated by some bounded operators. The conjugated elements themselves are 
of Hilbert-Schmidt class. From this we conclude that our condition is basis 
independent. We point out that some variants   
of this argument on Hilbert-Schmidt condition will be used over and over again. 
Same for $\theta$ except that  $\theta$ only has odd terms.
Notice that the second of these conditions is automatically true  since  the 
odd space is finite dimensional. In a more general case we will mention later on,
there will be extra convergence conditions on the odd generators.
In this setting $w$ is even and $\theta$ is odd.
For computations it is sometimes better to decompose a given matrix into 
its ordinary part and its nilpotent part, just like a super number 
being decomposed into an ordinary complex number plus the rest.
We use the terminology of deWitt and call it body and soul decomposition.
For example $w=w_B+w_S$ and $\theta=\theta_S$.
Let us define the restricted super-disc as the algebra of functions generated by  the 
above set of super-operators $Z$ with a further condition on $w$, 
\beq
    1-w_B^\dag w_B>0\quad 
.\eeq
Notice that we can interpret these to be 
 the elements which generate the $C^\infty$ functions on 
the superdisc. 
For later use we must  give a meaning to 
$Z^\dag Z$, so we define it to be the tensor product,
$Z^\dag Z=\pmatrix{w^\dag w & w^\dag \theta\cr \theta^\dag w& \theta^\dag \theta}$.
We do not demand any extra conditions on the $\theta$ variable. 
The inverse of $1-Z^\dag Z$ can be computed; we write 
\beq 
    (1-Z^\dag Z)^{-1}=1+Z^\dag(1-ZZ^\dag)^{-1}Z
\eeq
similarly for $w_B$ we have 
$(1-w_B w_B^\dag)^{-1}= 1+w_B(1-w_B^\dag w_B)^{-1}w_B^\dag$ and the 
operator on the right is well-defined due to positivity condition,
this means that the inverse on the left also exists.
Since we use  a  finite dimensional odd-space 
we can  define
\bea
   (1-ZZ^\dag)^{-1}&=&(1-w_Bw_B^\dag-w_B^\dag w_S-w_S^\dag w_B-w_S^\dag w_S-
\theta\theta^\dag)^{-1}\nn\cr
 &=&[1-(1-w_Bw_B^\dag)^{-1}(w_B^\dag w_S+w_S^\dag w_B+w_S^\dag w_S+ 
\theta\theta^\dag)]^{-1}(1-w_B w_B^\dag)^{-1}
,\eea
the first inverse in the last term can be expressed  via  a terminating expansion,
\bea
  [1&-&\!\!\!\!(1-w_Bw_B^\dag)^{-1}(w_B^\dag w_S+w_S^\dag w_B+w_S^\dag w_S+
\theta\theta^\dag)]^{-1}=\nn\cr
&=&1+(1-w_Bw_B^\dag)^{-1}(w_B^\dag w_S+w_S^\dag w_B+w_S^\dag w_S+\theta\theta^\dag)-\nn\cr
...&+&(-1)^{s-1}[(1-w_Bw_B^\dag)^{-1}(w_B^\dag w_S+
w_S^\dag w_B+w_S^\dag w_S+\theta\theta^\dag)]^s
\eea
where we assume that the degree of nilpotency of the 
supermatrix is  $s+1$.
We note that this is a general fact, if the body of a matrix is invertible then the 
matrix is invertible.
This series does not have to terminate in the infinite dimensional case,
so one has to impose the invertibility condition separately, or 
assume that the infinite formal expansion can be given a meaning( see the book 
by deWitt \cite{dewitt}).
The definition we propose later on may  result in a  
deviation from the Kostant-Berezin-Leites definition \cite{kostant, berezin1}.

There is a natural super-operator on the space $\Hi$ given with respect to the 
above direct sum as:
\beq
     J=\pmatrix{1&0&0\cr 0&-1&0\cr 0&0&-1}
.\eeq
Similar to  the finite dimensional case, we have an action of a 
certain super-pseudounitary group on the super-disc $\Di$.
Let us define the set of superoperators $g :\Hi\to \Hi$ with a
bounded inverse, such that they leave the operator $J$ invariant:
\beq
      gJg^\dag=J
.\eeq
Let us explicitly write this condition in a block decomposition:
\beq
     g=\pmatrix{A&B\cr C&D}
\eeq
and here 
$A:\Hi^e_-\to \Hi^e_-$, $B:\Hi^e_+|\Hi^o_+\to \Hi^e_-$,
$C:\Hi^e_- \to \Hi^e_+|\Hi^o_+$, finally 
$D: \Hi^e_+|\Hi^o_+\to \Hi^e_+|\Hi^o_+$.
This representation is better suited for our needs.
We have then
\beq
  AA^\dag-BB^\dag=1 \quad CA^\dag=DB^\dag \quad D D^\dag -C C^\dag=1
.\eeq
Using the invertibility we see that 
$g^\dag J g=J$ is also true, hence we get 
\beq
   A^\dag A-C^\dag C=1\quad A^\dag B=C^\dag D\quad D^\dag D-B^\dag B=1
.\eeq
 The first one means in terms of body and soul decomposition,
\beq
A_B A^\dag_B-B_B B_B^\dag =1 \quad A_S^\dag A_B+A_B^\dag A_S+A_S^\dag A_S+C^\dag_B C_S
+C^\dag_S C_B+C_S^\dag C_S=0
\eeq
and similarly for the others.
This means that the body parts of these matrices obey exactly the 
usual pseudounitary conditions hence we can do everything in the same way like
the non-supercase.
Among these set of opeators we pick the ones which satisfy a convergence condition,
written with respect to the direct sum decomposition $\Hi^e_-\oplus\Hi^e_+|\Hi^o_+$:
\beq
  g=\pmatrix{\BO&\ids&\ids\cr \ids &\BO&\BO\cr \ids&\BO&\BO}
,\eeq
and these conditions are imposed on the components of each term,
i.e. if we expand the upper corner, $\beta=\beta_a \xi^a+\beta_{a_1a_2a_3}\xi^{a_1}\xi^{a_2}
\xi^{a_3}+...$, each term belongs to $\ids$ and similarly for the other parts.
We may also  economically express these  in the form $[J,g] \in \ids$, with the 
above interpretation for the ideal.
Therefore we can summarize the above set of operators in the form of 
a group: 
\beq 
  U_1(\Hi^e_-,\Hi^e_+|\Hi^o_+)=\{ g| g^{-1} \ {\rm exists,} \ [J,g] \in 
\ids \quad {\rm and }\ gJg^\dag=J\}
,\eeq
where the ideal condition refers to our convention.
The main point is to show that the convergence conditions hold after the multiplication.
This follows the same line of arguments as before, if one writes explicitly the components,
we see that each one is a finite sum of Hilbert-Schmidt operators.
We leave it to the reader to check the details.
This group is one possible super version of the pseudounitary group.
We refer to this set  the restricted super-pseudounitary group.

Just like  the classical case, the restricted super-pseudounitary group has an 
action on the super-disc $\Di$.
This action is written in the super-operator language 
 exactly as in the classical case:
\beq
      Z\mapsto (AZ+B)(CZ+D)^{-1}
,\eeq
where we use 
\beq
     g=\pmatrix{A&B\cr C&D}
\eeq
We need to clarify the action of $C$,
if we denote $C$ as $\pmatrix{c\cr \gamma}$,
\beq
CZ=C\otimes Z=\pmatrix{cw& c\theta\cr \gamma w &\gamma \theta},
\eeq
which shows that the action is well-defined and the resulting 
operator goes from 
$\Hi^e_+|\Hi^o_+$ to $\Hi^e_+|\Hi^o_+$, thus we can add $D$ to this.
Let us note that the inverse on the right exists,
this is because the even part has an inverse and we can define 
the inverse by a terminating expansion.
Just for an illustration we give the explicit version,
the reader who is familiar with this kind of manipulations is advised 
to skip this part:
We would like to show that  $CZ+D$ has an inverse.
We know that $D^{-1}$ is well-defined,
hence it is better to look at 
$D^{-1}CZ+1$.
We use the following formula for the inverse of a supermatrix:
\beq
\pmatrix{ {\tilde A}& {\tilde B}\cr {\tilde C}& {\tilde D}}^{-1}=
\pmatrix{({\tilde A}-{\tilde B}{\tilde D}^{-1}{\tilde C})^{-1}&-{\tilde A}^{-1}{\tilde B}
({\tilde D}-{\tilde C}{\tilde A}^{-1}{\tilde B})^{-1}\cr
   -{\tilde D}^{-1}{\tilde C}({\tilde A}-{\tilde B}{\tilde D}^{-1}{\tilde C})^{-1} 
&({\tilde D}-{\tilde C}{\tilde A}^{-1}{\tilde B})^{-1}}
\eeq
This can be written in the following form:
\beq
\pmatrix{ 1+(d_{11}-\delta_{12}d_{22}^{-1}\delta_{21})^{-1}cw-d_{11}^{-1}\delta_{12}
(d_{22}-\delta_{21}d_{11}^{-1}\delta_{12})^{-1}\gamma w& ***\cr
***& \!\!\!\!\!\!\!\!\!\!\!\!\!\!\!\!\!\!\!\!\!\!\!\!\!\!\!\!\!\!\!\!\!\!\!\!\!\!
\!\!\!\!\!\!\!\!\!\!\!\!\!\!\!
1-d_{22}^{-1}\delta_{21}(d_{11}-\delta_{12}d_{22}^{-1}\delta_{21})^{-1}c\theta+
(d_{22}-\delta_{21}d_{11}^{-1}\delta_{12})^{-1}\gamma \theta}
.\eeq
To prove the invertibility we do not need the explicit forms of the 
off-diagonal components, this is why they are not shown in the above matrix.
The lower diagonal block is invertible,
due to the nilpotency of the part added to $1$.
Hence we need to check only the upper diagonal block(actually, this is a general 
result).
To do this we recall that the super-pseudounitarity means,
$DD^\dag=1+CC^\dag$, written in terms of components, the upper block gives us
$d_{11}d_{11}^\dag=cc^\dag+1-\delta_{12}\delta_{12}^\dag$.
This means that we can define an inverse  square root of the above matrix;
for this we use the following integral representation,
\beq
   T^{-1/2}={1\over \pi}\int_0^\infty {d\lambda\over \lambda^{1/2}}(T+\lambda I)^{-1}
\eeq
This formula is used for a positive operator and can be extended to the 
super-case when the body of the super-matrix  
 is positive(this is the case for us as we will see shortly).
As a result we get,
\bea
 (d_{11}d_{11}^\dag)^{-1/2}&=&{1\over \pi}\int_0^\infty {d\lambda\over \lambda^{1/2}}
\Big[(c_Bc_B^\dag+\lambda 1+1)^{-1}+(c_Bc_B^\dag+\lambda 1+1)^{-1}f_S
(c_Bc_B^\dag+\lambda 1+1)^{-1}+...\cr\nn
&\ & -(-1)^r((c_Bc_B^\dag+\lambda 1+1)^{-1}f_S)^r
(c_Bc_B^\dag+\lambda 1+1)^{-1}\Big]
.\eea
here  we use
$f_S=c_Bc_S^\dag+c_Sc_B^\dag+c_Sc_S^\dag+\delta_{12}\delta_{12}^\dag$
which is a nilpotent matrix and we assumed that it has degree $r+1$.
The first term is the usual term $(1+c_Bc_B^\dag)^{-1/2}$, the others are 
nilpotent contributions.
Hence when we write this first diagonal block in this form,
\beq
  1+u(1+c_Bc_B^\dag)^{-{1\over 2}}c_Bw_B+{\rm nilpotent \quad parts}
\eeq
where $u$ is a unitary piece that we cannot determine--stripped off from its 
possible nilpotent part.
It is enough to show that this leading part is invertible, but that is 
the same as in non-super case:
$||w_B^\dag c_B^\dag(1+c_Bc_B^\dag)^{-1/2}u^\dag u
(1+c_Bc_B^\dag)^{-1/2}c_Bw_B||\leq||w_B^\dag w_B|| <1$,
this implies that the series expansion will converge and we have an invertible 
element.
This concludes our demonstration.
Of course we have done more than just showing that the inverse is well-defined,
we also got an expasion of the inverse, which is useful to show the 
convergence condition in the infinite dimensional case.
There is a simpler way to show the invertibility, which we repeat here for clarity,
\beq 
(CZ+D)=\pmatrix{cw+d_{11} & c\theta+\delta_{12}\cr \gamma w+\delta_{21} & \gamma \theta+d_{22}}
,\eeq
and as we have observed for invertibility it is enough to know the invertibility 
of the body parts,
we have $d_{11} d_{11}^\dag=cc^\dag +1-\delta_{12}\delta_{12}^\dag $,
the body parts satisfy $(d_{11})_B(d_{11}^\dag)_B=1+c_Bc_B^\dag$, and using the same argument 
as before this implies 
that the body  is invertible(we already know $d_{22}$ is invertible).
 
One can check that the resulting operator 
$Z'$ is an element of the super-disc.
We briefly indicate how this is done:
The convergence conditions are easy since we have 
$Z\in\ids$ and $B\in \ids$.
If we want to show that the resulting operator satisfies 
$1-w^{'\dag}_B w'_B>0$, we look at 
$\Big ((AZ+B)(CZ+D)^{-1}\Big )_B$,
this comes from $w'$,
\beq
   w'= (Aw+b)(cw+d_{11}-(c\theta+\delta_{12})(\gamma \theta+d_{22})^{-1}
(\gamma w+\delta_{21}))^{-1}
,\eeq
which has body part:
\beq
   w'_B=(A_Bw_B+b_B)(c_Bw_B+(d_{11})_B)^{-1}
.\eeq
We have from the pseudo-unitarity  conditions,
$A_B A_B^\dag-b_B b_B^\dag=1$, $c_B A_B^\dag =(d_{11})_B b^\dag_B$,
$(d_{11})_B(d_{11}^\dag)_B-c_B c_B^\dag=1$.
But these are exactly the conditions for the ordinary pseudo-unitary group
$U_1(\Hi_-^e,\Hi_+^e)$, hence the positivity condition follows as in the 
ordinary case.
Of course the point is to show that the action is transitive,
and hence to prove that the super-disc is a homogeneous manifold.
Let us go over this point as well using similar techniques to the above proof.
To prove this it is enough to show that the action is 
transitive over the generating set of elements for the $C^\infty(D^I_1)$ we introduced,
$Z=[w\quad \theta]$.(Notice that a super manifold is really defined through the algebra 
of functions living on it).
Let us show that we can obtain all the generators  starting from 
$Z=0$ using the group action.
Recall that the pseudo-unitarity imposes the following conditions,
\beq
  AA^\dag-BB^\dag=1 \quad CA^\dag=DB^\dag \quad D^\dag D-B^\dag B=1
,\eeq
the last one uses the opposite multiplication.
For any $Z=BD^{-1}$, if we insert this into the last one 
we  see that 
$D=(1-Z^\dag Z)^{-1/2}U$, where $U$ is an arbitrary 
super-unitary element acting on the same space, is a solution.
Later on we will prove that this square root makes sense 
and the body belongs to the desired class, but first we will present the formal 
solution in the super-matrix form:
\beq
 g=\pmatrix{(1-ZZ^\dag)^{-1/2}V & Z(1-Z^\dag Z)^{-1/2}U\cr 
       (1-Z^\dag Z)^{-1}Z^\dag(1-ZZ^\dag)^{1/2}V& (1-Z^\dag Z)^{-1/2}U}
.\eeq
where $V\in U(\Hi_-^e)$ and $U\in U(\Hi_+^e|\Hi_+^o)$.
In fact this shows the ambiguity in the solution to be exactly 
the subset  we mode out with.
Let us prove the claim using the integral form of the square root of the 
matrix, we begin with $A$,
\bea
   (1-ZZ^\dag)^{-1/2}&=&{1\over \pi}\int_0^\infty 
{d\lambda\over \lambda^{1/2}}(\lambda 1+1-ww^\dag -\theta\theta^\dag)^{-1}\nn\cr
&=&(1-w_Bw_B^\dag)^{-1/2}+{1\over \pi}\int_0^\infty {d\lambda\over \lambda^{1/2}}
(\lambda 1+1-w_Bw_B^\dag)^{-1} f_S (\lambda 1+1-w_Bw_B^\dag)^{-1}+...\nn\cr
&-&(-1)^r{1\over \pi}\int_0^\infty {d\lambda \over \lambda^{1/2}}(\lambda 1+1-w_Bw_B^\dag)^{-1}
[f_S(\lambda 1+1 -w_Bw_B^\dag)^{-1}]^r
,\eea
where $f_S=w_Sw_B^\dag+w_Bw_S^\dag+w_Sw_S^\dag+\theta \theta^\dag$.
Notice that everything is well-defined here.
Let us now indicate  that $D$ is well-defined, we do this for the upper 
corner only,\footnote{Just for fun, we 
suggest the reader to show the following identity, which gives an alternative proof of the
existence,
\beq
  (1-Z^\dag Z)^{-1/2}=1+Z^\dag((1-ZZ^\dag)^{-1/2}-{1\over 2}\int_0^1 dt 
(1-tZZ^\dag)^{-1/2})Z
\eeq}
\beq
   D=(1-Z^\dag Z)^{-1/2}={1\over \pi}\int_0^\infty {d\lambda \over \lambda^{1/2}}
\pmatrix{\lambda 1+1-w^\dag w & -w^\dag \theta\cr
 -\theta^\dag w& \lambda 1 + 1-\theta^\dag \theta}^{-1}
\eeq
As a result of this expression we se that all the elements are well-defined and 
belong to the correct classes, in fact  we can write the expansion for 
$d_{11}$, 
\bea 
   d_{11}&=&(1-w_B^\dag w_B)^{-1/2}+{1\over \pi}\int_0^\infty {d\lambda\over \lambda^{1/2}}
(\lambda 1+1-w_B^\dag w_B)^{-1}\nn\cr
&\times&(w_B^\dag w_S+w_S^\dag w_B+w^\dag_Sw_S+w^\dag 
\theta(\lambda 1+1-\theta^\dag \theta)^{-1}\theta^\dag w)(\lambda 1+1-w_B^\dag w_B)^{-1}+....
\eea
where the series terminates.
we see that everything is well-defined. One can see that the rest of it can be done in a 
simple way since the expressions for  $B,C$ have explicit multiplicative factors of 
$Z$ which is in the Hilbert-Schmidt class, so we skip the details for brevity.
Let us  also check again  the stability subgroup of $Z=0$ is given by 
$U(\Hi^e_-)\times U(\Hi^e_+|\Hi^o_+)$.
For $Z=0$, $Z'=BD^{-1}$, if we set this to zero,
since $D $ is invertible, we get 
$B=0$.
>From the invariance of $J$ we get 
$AA^\dag=1$, and this together with 
$AC^\dag-BD^\dag=0$ implies $C=0$.
The result of this is the diffeomorphim we are after:
\beq
    \Di=U_1(\Hi^e_-,\Hi^e_+|\Hi^o_+)/U(\Hi^e_-)\times U(\Hi^e_+|\Hi^o_+)
.\eeq
We emphasize that the explicit coordinate $Z$ shows that this is
a super-complex manifold, the group action point of view instead 
shows that this space is a super homogeneous space.

\section{Supersymplectic Structure}

In this section we will discuss the classical mechanics on this super disc.
There is a natural supersymplectic structure, it is homogeneous and 
further more it is K\"ahler. This is a natural choice from the point of view of geometry and 
as we will see it also provides us a natural method of quantization, which is an extension of 
the Bargmann representation to this case\cite{2dgeo}.
The analysis of symmetric domains and the use of Toeplitz operators in the quantization 
problem is thoroughly discussed in the book by Upmeier \cite{upmeier}.
We also recommend the articles by Borthwick et al\cite{borthwick}.

It will be  simpler to use the  
following super operator to show that the disc is a supersymplectic space;
\beq
          \Phi=-1+2\pmatrix{(1-ZZ^\dag)^{-1}&-(1-ZZ^\dag)^{-1}Z\cr
                Z^\dag (1-ZZ^\dag)^{-1}&-Z^\dag (1-ZZ^\dag)^{-1}Z}
,\eeq
Notice that this operator is well-defined on $\Hi_-^e\oplus\Hi_+^e|\Hi_+^o$.
The reader can check that 
\beq
    \Phi^2=1 \quad \ J\Phi^\dag J=\Phi
.\eeq
An important point is that the action of the group on $Z$ becomes very simple on $\Phi$,
$Z\to g\circ Z$ induces $\Phi \to g\Phi g^{-1}$(see Appendix).
$Z=0$ corresponds to $\Phi=J$, and we can check that 
$\Phi(Z)=g(Z)Jg(Z)^{-1}$(see Appendix).
We may define a symplectic form on $D^I_1$ using $\Phi$;
formally,
\beq
     \Omega={i\over 4} \Str \Phi d\Phi\wedge d\Phi
.\eeq
What we mean by this two form is that if we take two vector fields $V_u, V_v$,
which are generated by 
the action of the super-Lie group, we get a number: 
\beq
    \Omega(V_u,V_v)={i\over 8}\Str J[[J,g^{-1}ug]_s,[J,g^{-1}vg]_s]_s
.\eeq
Using the above formal expression, we see that $\Omega$ is closed and furthermore 
it is homogeneous. This easily follows from the transformation of 
$\Phi$ under the group action. One can actually see this by looking at the explicit form 
of it.
The nondegenracy and super-K\"ahler structures are best understood around $J$,
then we use the homogeneity to distribute this form over all the manifold.
When we restrict ourselves to the point $J$:
\beq
    \Omega|_{Z=0}=i\Str\pmatrix{-dZ\wedge dZ^\dag& 0\cr 0& dZ^\dag \wedge dZ} 
,\eeq
here the two wedge products have different meanings:
$dZ\wedge dZ^\dag=dw\wedge dw^\dag +d\theta\wedge d\theta$, and 
$dZ^\dag\wedge dZ=\pmatrix{dw^\dag \wedge dw& dw^\dag \wedge d\theta\cr d\theta^\dag\wedge dw&
 d\theta^\dag\wedge d\theta}$. Hence we can rewrite this expression as,
\beq
    \Omega|_{Z=0}=i\Str \pmatrix{-dw\wedge dw^\dag-d\theta\wedge d\theta^\dag &0\cr
      0& \pmatrix{dw^\dag \wedge dw& dw^\dag \wedge d\theta\cr d\theta^\dag\wedge dw&
 d\theta^\dag\wedge d\theta}}
.\eeq
By expanding the trace,
\bea
   \Omega|_{Z=0}&=&i[-\Tr dw\wedge dw^\dag-\Tr d\theta\wedge d\theta^\dag+
\Str\pmatrix{dw^\dag \wedge dw& dw^\dag \wedge d\theta\cr d\theta^\dag\wedge dw&
 d\theta^\dag\wedge d\theta}]\cr\nn
&=&i[-\Tr dw\wedge dw^\dag-\Tr d\theta\wedge d\theta^\dag+
\Tr dw^\dag \wedge dw-\Tr d\theta^\dag \wedge d\theta]\cr\nn
&=&-2i\Tr dw\wedge dw^\dag -2i \Tr d\theta\wedge d\theta^\dag=-2i\Tr dZ\wedge dZ^\dag.
\eea
(This incidentally  shows that the form is super-K\"ahler\cite{gradechi1, gradechi2,manin}).
By contracting this with two vector fields at the origin, we get 
\beq
     \Omega(V_u,V_v)|_{Z=0}=-2i[\Tr (b_1b_2^\dag -b_2 b_1^\dag)+\Tr(\beta_1\beta_2^\dag+
\beta_1\beta_2^\dag)]
.\eeq
Notice that we use the vector $[b\ \beta]$ for the component $u_{12}$ of 
the Lie algebra element (this could be somewhat confusing but we try to avoid the prolifiration
of indices).  
Using the above form it is possible to prove the nondegeneracy, this is given in the 
Appendix.
The symplectic form above provides us with a Poisson structure,

 One can define  classical dynamics on this superspace,
given an even Hamiltonian, a physical observable, $H$.
The time evolution of any observable $O$ is given by
\beq 
      {\partial O\over \partial t}=\{H,O\}_s
\eeq

One can naturally ask if there are moment maps which generate the goup action.
It is not possible to use 
$F_u={i\over 2}\Str \Phi u$ due to divergence of the trace,
 but it is possible to do a vacuum subtraction and 
get a convergent one.
To prove this we use a rearrangement of the formula   for $\Phi$:
\beq 
    \Phi(Z)=J+\pmatrix{2Z(1-Z^\dag Z)^{-1}Z^\dag &-2(1-ZZ^\dag)^{-1}Z\cr
                 2Z^\dag (1-ZZ^\dag)^{-1} &-2Z^\dag(1-ZZ^\dag)^{-1}Z}
,\eeq
If we look at now the difference,
$\Phi(Z)-J$ the last part remains.
The diagonal parts of this operator are better behaved than the off-diagonal parts,
$Z(1-Z^\dag Z)^{-1} Z^\dag \in \idf$ in our sense as one can see.
and similarly for the other one.
The off-diagonal parts are actually in $\ids$.
So when we look at $(\Phi(Z)-J)u$, we see that,
\beq
    \pmatrix{ \idf&\ids&\ids \cr \ids &\idf &\idf\cr \ids&\idf&\idf}
\pmatrix{{\cal B}&\ids&\ids \cr \ids &{\cal B} &{\cal B}\cr \ids &{\cal B}&{\cal B}}
=\pmatrix{\idf&\ids &\ids\cr \ids&\idf &\idf \cr \ids &\idf &\idf }
.\eeq
Hence a conditional trace exists:
if we throw away the nontrace parts,
$\Str_J(\Phi(Z)-J)u={1\over 2}\Str[(\Phi(Z)-J)u+J(\Phi(Z)-J)uJ]$ is actually convergent.
We see that this is very similar to the ordinary disc case in \cite{2dgeo}.

A general discussion shows that we get  
a Poisson realization of the super-Lie algebra through the 
moment maps:
\beq
      \{ F_u, F_v\}_s=F_{[u,v]_s}+\Sigma_s(u,v)
\eeq
It is possible to find this central term by evaluating everything 
at the origin, $\Phi=J$:
 \bea
    \Sigma_s(u,v)&=&{i\over 8} \Str
 \pmatrix{1&0&0\cr0&-1&0\cr 0& 0& -1}\Big[\Big[\pmatrix{1&0&0\cr0&-1&0\cr 0& 0& -1},u\Big],
  \Big[\pmatrix{1&0&0\cr0&-1&0\cr 0& 0& -1}, v\Big]\Big]_s\cr\nn
  &=&{i \over 2} \Str_J [ J, u ]v
,\eea
It is interesting to write down the central term explicitly:
\bea
  \Sigma_s(u,v)&=&-2i 
\Str\pmatrix{b_1b_2^\dag +\beta_1\beta_2^\dag & *\cr *&\pmatrix{b_1^\dag b_2&
 b_1^\dag \beta_2\cr \beta_1^\dag b_2&\beta_1^\dag \beta_2}}\nn\cr
 &=&-2i(\Tr(b_1b_2^\dag-b_2b_1^\dag)+\Tr(\beta_1\beta_2^\dag)-\Tr(\beta_1^\dag \beta_2))\nn\cr
  &=&-2i((\Tr(b_1b_2^\dag-b_2b_1^\dag)+\Tr(\beta_1\beta_2^\dag+\beta_1 \beta_2^\dag))\nn\cr
\eea
and we see that at each step the diagonals are in ${\cal I}_1$ and hence the traces are all 
well-defined. This is equal to the symplectic form at the origin we computed before 
using the explicit coordinate $Z$ as it should be.
This type of central term is expected when there are bosons and fermions mixed.
An interesting discussion of such central extensions from the Fock space point of view is
given in \cite{grosse}.In \cite{ekstrand1} ${\bf Z}_2$ graded Schwinger terms for
neutral particles are worked out,
in \cite{ekstrand2} current super-algebras are studied, providing
a generalization of Mickelsson-Rajeev cocyle\cite{mickelsson}. The use of pseudodifferential
operators in this reference
we believe is better motivated in these higher dimensional cases.
There should be a similar extension of our results using this restricted class of operators.

We are therefore equipped with a powerful geometric setting to develop 
our geometric quantization program.
 
\section{Geometric Quantization}

Our presentation here will be somewhat more concise, 
most of the computations can be done similar to our previous work, excapt one has to watch
for the signs.
The technical details and explanation of the main ideas are already given in
\cite{kostant}, we recommend  the  examples in  \cite{gradechi1, gradechi2}, and one can 
read a more general program in
\cite{vaisman}(we believe it is interesting  to follow the philosophy of the last reference).

We can follow exactly the same steps in  \cite{2dgeo} introduce a prequantization line bundle
(for ordinary geometric quantization we refer to \cite{hurt, kirillov, kirillov2, woodhouse}),
and  we introduce a super-one form on this bundle:
\beq 
  \Theta_s= {1\over \hbar}( \Str(1-Z^\dag Z)^{-1}dZ^\dag Z-\Str (1-Z^\dag Z)^{-1}Z^\dag dZ)
.\eeq
This is use to define the covariant derivate as in the 
nonsuper case\footnote{ strictly speaking in the model of super-sections this 
acts on the a prolongation, $\Gamma(M, \wedge({\bf C}^N)\otimes K)$ where $K$ is a 
prequantum complex line bundle on the base $M$.},
\beq
     \nabla_V={\cal L}_V^s+\Theta_s(V)
,\eeq
where we used a superscript to denote the super-Lie derivative.
For any given super-function,
we have the vector field generated from the symplectic form,
\beq
    \Omega(V_f,*)=-df
\eeq
using this vector field a prequantization operator is obtained,
\beq
   \tilde f=-i\hbar \nabla_{V_f}+ f
\eeq
This gives us a representation of the Poisson brackets:
\beq
    \widetilde{\{f, g\}_s}=-i\hbar[ \tilde f, \tilde g]_s
.\eeq

As in the ordinary case, we need to restrict the prequantum Hilbert space,
since the prequantization map does not lead to an irreducible representation.
We will choose super-holomorphic functions,
\beq
    \nabla_{Z^\dag} \psi(Z,Z^\dag)=0
.\eeq
The super analysis is designed to provide a complete analogy to the usual 
analysis, hence most of what we said follows from a routine yet long(and care 
required due to signs) computations.

We can solve for this  holomorphicity condition as in the ordinary case:
\beq
\nabla_{Z^\dag}\psi=0, \quad \psi(Z,Z^\dag)={\rm sdet}^{1\over \hbar}(1-Z^\dag Z)\Psi(Z)
\eeq
where $\Psi(Z)$ denotes a superholomorphic function on the disc.
We define the superdeterminant(or Berezinian) as 
\beq 
      {\rm sdet} \pmatrix{\tilde A&\tilde B\cr \tilde C&\tilde D}=
{\rm det}(\tilde A-\tilde B\tilde D^{-1}\tilde C)({\rm det} \tilde D)^{-1} 
,\eeq
where the operator is written  according to the even and odd decomposition of the
super Hilbert space. The infinite dimensionality of the underlying space  requires  the
full operator to be of the form $1+{\cal I}_1$, otherwise one has to use a 
conditional determinant.
The resulting operators for the moment maps acting on holomorphic sections will be exactly 
the same as in the ordinary case,
\beq
       \hat F_u\Psi(Z)=-i\hbar[{\cal L}_{V_u}^s-{1\over \hbar}\Str( u_{21} Z)]\Psi(Z)
\eeq
where we have used the same letters to denote the components of the  
Lie algebra elements, $u=\pmatrix{ u_{11} &u_{12} \cr u_{21}&u_{22}}$,
not to bring new notation.
Holomorphicity is clearly preserved and these are the 
correct operators to start a quantization program.

These moment maps 
 can be integrated 
to a representation of a central extension of the  super-pseudounitary group:
\beq
    \rho(g^{-1})\Psi(Z)={\rm sdet}^{-{1\over \hbar}}(D^{-1}CZ+1)\Psi((AZ+B)(CZ+D)^{-1})
\eeq
This is a well-defined representation, let us see that the determinant exits:
\beq
   \pmatrix{ (d^{-1})_{11}& (\delta^{-1})_{12}\cr (\delta^{-1})_{21}& (d^{-1})_{22}}
   \pmatrix{ cw& c\theta\cr \gamma w & \gamma \theta}=\pmatrix{{\cal B} & {\cal B} \cr 
{\cal B} & {\cal B} } 
\pmatrix{{\cal I}_1&{\cal I}_1\cr
{\cal I}_1 & {\cal I}_1}=\pmatrix{{\cal I}_1&{\cal I}_1\cr
{\cal I}_1 & {\cal I}_1}
.\eeq
This shows that the determinant is absolutely convergent--independent of the basis chosen--. 
The central term of the representation  is given by,
\beq
      c_S(g_1,g_2)={\rm sdet}^{1\over \hbar}[(D_1D_2)^{-1}C_1B_2+1]
,\eeq
derivation  of this supercentral term does not present any more difficulties then the 
ordinary case(see the appendix of \cite{teo}), convergence issue 
follows the same lines as the above one,  we leave the details to the reader.

\section{Infinite dimensional case}

We will propose a way of extending our results when 
${\cal H}^o$ is infinite dimensional. In this section we will not repeat the previous 
arguments, since some of them are direct generalizations and some of them require a much 
deeper study. We plan to come back to those issues in another publication, so 
in this section we only give a sketch of ideas.
While we were working on this problem, we became aware of a  rather similar
set of ideas by Schmidt in \cite{schmidt}.  Which set of ideas 
are more appropriate for our problem is not so clear to us at this moment, so we follow 
our point of view, we plan to take a more detail study of all these issues in the future.

First we change our notion of a super-number:
\beq 
     z=z_B+\sum_{N=0}^{\infty} \sum_{a_1<a_2<...<a_N} z_{a_1a_2...a_N}\xi^{a_1}
\xi^{a_2}...\xi^{a_N},
\eeq
where  
we assume that the sums have square integrable coefficients
$\sum_N^\infty \sum_{a_1<a_2<...<a_N} |z_{a_1a_2...a_N}|^2<\infty$.
This makes the product of two super numbers well-defined, hence behaves much better 
than the formal sums, it is physically  more transparent as well.
The product becomes
\beq 
    (z t)_{a_1a_2...a_N}=\sum_n z_{(a_1a_2...a_{n}}t_{a_{n+1}...a_N)}
,\eeq
here $(...)$ denotes an appropriate symmetrization  of the indices,
due to the ordering of the generators(keeping the previous ordering in mind).
>From a more abstract point of view 
when we look at the algebra of smooth  functions  on this flat space 
we get 
$C^\infty(F) \approx  { {\oplus}}_{l^2} \wedge^k {\cal H} $
and this is what defines the cartesian product of supernumbers.
We will naturally represent the rigth hand side 
as the naive Fock space of the Hilbert space:
${\cal F} ({\cal H})={ {\oplus}}_{l^2} \wedge^k {\cal H} $.(This 
is not the Fock space corresponding to the Dirac sea, it is the naive one).
We look at again the mtrix algebra modeled on these super-numbers,
they will be transformations from 
$Z: {\cal H}^e_+ |  {\cal H}^o_+ \to {\cal H}^e_-$ written explicitly,
   $ Z=Z_B+\sum_N \sum_{a_1<a_2<...<a_N} Z_{a_1a_2...a_N}$, 
matrix coefficients satifying,
\beq
     \sum_N \sum_{a_1<a_2<...<a_N} ||Z_{a_1a_2...a_N}||_2 ^2<\infty
\eeq
where $||*||_2$ denotes the norm in the Hilbert-Schmidt ideal.
This implies that  we have a space of matrices which 
is modeled on 
${\cal I }_2 \otimes {\cal F} ({\cal H})$.
We may use the above convergence condition  to get an inner product:
\beq 
< Z, W >=\sum_N \sum_{a_1<a_2<...<a_N} \Tr Z_{a_1a_2...a_N} ^\dag W_{a_1...a_N}
.\eeq
We note that this abstract space is still a Hilbert space with
the above inner product, and indeed that will equip  us with 
all the luxuries of  Hilbert spaces.
We can prove by using standard techniques that the product of two such 
matrices, $ZW$ is still in the above  class,
i.e.
\beq
    \sum_N \sum_{a_1<a_2<...<a_N} ||(ZW)_{a_1a_2...a_N}||^2_2=
   \sum_N \sum_{a_1<a_2<...<a_N} ||\sum_n Z_{(a_1a_2...a_n}
W_{a_{n+1}...a_N)}||^2_2<\infty
.\eeq
There is the same type possible reorderings of  the indices  in this expression.
The rest will follow exactly the same lines as before, the convergence conditions should be 
checked much more carefully this time.

The disc is defined as 
$ 1-w_B^\dag w_B>0$, and
$Z=[w\  \theta]$.
where each one of these super-matrices satisfy the above condition for 
being in  ${\cal I}_2$.
We can define the same symplectic form, 
\beq
     \Omega(V_u,V_v)={i \over 8}\Str J [[J, g^{-1} u g]_s, [J, g^{-1} v g]_s]_s
\eeq
here each term is in ${\cal I}_2$.

 The rest of the arguments apart from the convergence issues are exactly the same, so we leave
the details to a future work.

\section{Acknowledgements} The author would like to thank M. Arik, K. Gawedzki, P. Guha, 
G. Grahovski, J. Gracia-Bondia,
A. Konechny, I. Mladenov, J. Mickelsson, A. Nersessian, R. Nest, 
S. G. Rajeev, C. Saclioglu, S. Scott,  M. Walze for 
discussions and several useful comments. The author also greatfully acknowledges 
the kind invitation from IHES while this work is in progress.

\section{Appendix}
For comleteness we define here $A^\alpha$ where $0<\alpha<1$ and 
the body of the super operator is positive,
\beq 
        A^\alpha={\sin \pi \alpha\over \pi}\int_0^\infty {d\lambda\over \lambda^{1-\alpha}}
(\lambda 1+A)^{-1}
.\eeq
The advantage of this expression is that we may actually expand the inverse 
and obtain a series for the super operator.
Note that there is no simple recursive process when $\alpha $ is not a rational number.

In this part we will give a proof of the following transformation rule:
$Z \mapsto g\circ Z$ implies $\Phi \mapsto g \Phi g^{-1}$.
First we note that 
when $Z\mapsto (AZ+B)(CZ+D)^{-1}$, we have 
$(1-Z^\dag Z)^{-1} \mapsto (CZ+D)(1-Z^\dag Z)^{-1}(CZ+D)^{-1}$.
Next we rewrite $\Phi(Z)$: 
\beq
\Phi=-1+2\pmatrix{K^{-1} &-K^{-1}Z\cr Z^\dag K^{-1} &-Z^\dag K^{-1}Z}=1+
 2\pmatrix{Z^\dag S^{-1} Z & -ZS^{-1}\cr S^{-1}Z^\dag & -S^{-1}}
,\eeq
where $K=(1-ZZ^\dag)$ and $S=(1-Z^\dag Z)$.
Using the above observation we see that 
\beq
  \Phi(g\circ Z)=1+2\pmatrix{(AZ+B)S^{-1}(AZ+B)^\dag & -(AZ+B)S^{-1}(CZ+D)^\dag\cr
                             (CZ+D)S^{-1}(AZ+B)^\dag& -(CZ+D)S^{-1}(CZ+D)^\dag}
.\eeq

One can see that the above expression can be written as:
\beq
 \Phi(g\circ Z)=1+2\pmatrix{A&B\cr C&D}\pmatrix{Z^\dag S^{-1}Z&-ZS^{-1}\cr 
                        S^{-1}Z^\dag & -S^{-1}}\pmatrix{A^\dag & -C^\dag \cr 
                                                       -B^\dag & D^\dag }
,\eeq
this is precisely what we claimed.
Next point to check is $\Phi(Z)=g(Z)Jg(Z)^{-1}=g(Z)g(Z)^\dag J$:
\bea
    g(Z)g(Z)^\dag&=&\pmatrix{K^{-1/2}&ZS^{-1/2}\cr S^{-1/2}Z^\dag K^{1/2}& S^{-1/2}}
          \pmatrix{ K^{-1/2} & K^{1/2}ZS^{-1}\cr S^{-1/2}Z^\dag & S^{-1/2}}\cr\nn
         &=&\pmatrix{K^{-1}+ZS^{-1}Z^\dag+1-1&2ZS^{-1}\cr 2S^{-1}Z^\dag &
             S^{-1}Z^\dag KZS^{-1}+S^{-1}}
.\eea
Multiply this with $J=\pmatrix{1&0\cr 0&-1}$,
then in  the last line use $-(S^{-1}Z^\dag KZS^{-1}+S^{-1})=-(S^{-1}Z^\dag KK^{-1}Z+S^{-1})$,
which gives $-(S^{-1}Z^\dag Z+S^{-1})=S^{-1}(1-Z^\dag Z)-S^{-1}-S^{-1}$.
This gives $1-2S^{-1}$, then the result follows.

We will prove the non-degeneracy of the 
super-two form;
\beq 
       \Omega(V_u,V_v)|_{Z=0}=-2i\Tr(b_1b_2^\dag-b_2 b_1^\dag)-2i\Tr(\beta_1\beta^\dag_2+
   \beta_2\beta_1^\dag).
\eeq
Here we write $V_u(Z)=u_{11}Z-Zu_{22}-Zu_{21}Z+u_{21}$, and similarly for the 
$V_u(Z^\dag)$.
Furthermore we write for $u_{12}=[b\ \beta]$, hoping that the use of the same letters for 
the Lie algebra elements will not cause any confusion. 
Let us expand each term as a super-matrix(ignoring the multiplicative factor 
$-2i$),
\bea
   &\  & \Tr ((b_1)_B(b_2)_B^\dag-(b_2)_B(b_1)_B^\dag)=0\nn\cr
   &\  & \Tr ((b_1)_{a_1a_2}(b_2)_B^\dag-(b_2)_{a_1a_2}(b_1)_B^\dag)\xi^{a_1}
 \xi^{a_2}=0\nn\cr
   &\  & \Tr((b_1)_B(b_2)_{a_1a_2}^\dag -(b_2)_B(b_1)^\dag_{a_1a_2})\xi^{*a_1}\xi^{*a_2}=0\nn
\cr
   &\  &\Tr((b_1)_{a_1a_2a_3a_4}(b_2)_B^\dag \xi^{a_1}\xi^{a_2}\xi^{a_3}\xi^{a_4}
+(b_1)_{a_1a_2}(b_2)_{a_3a_4}^\dag\xi^{a_1}\xi^{a_2}\xi^{*a_3}\xi^{*a_4}
 +(b_1)_B(b_2)_{a_1a_2a_3a_4}^\dag\xi^{*a_1}\xi^{*a_2}\xi^{*a_3}\xi^{*a_4})\nn\cr
\!\!\!\!\!\!\!\!\!\!\!&-  & \Tr((b_2)_{a_1a_2}(b_1)^\dag_{a_3a_4}
\xi^{a_1}\xi^{a_2}\xi^{*a_3}\xi^{*a_4}
    +(b_2)_B(b_1)_{a_1a_2a_3a_4}^\dag\xi^{*a_1}\xi^{*a_2}\xi^{*a_3}\xi^{*a_4}\nn\cr
&+&(b_2)_{a_1a_2a_3a_4}(b_1)^\dag_B\xi^{a_1}\xi^{a_2}\xi^{a_3}\xi^{a_4})=0 \nn\cr
 &\ & ....    
\eea
where the dots refer to the continuation of this expansion. 
>From these relations we conclude that  
an iterative process gives us the required nondegeneracy.

\end{document}